\pdfoutput=1
\documentclass[11pt,twoside,a4paper,cmspaper,final,collab]{cms-tdr}
\def\svnVersion{3d25d2f}\def\svnDate{2022/09/12}\def\cmsCernNoTag{CERN-EP-2021-225}\def\cmsCernDate{\today}\def\cmsMessage{Published in Physics Letters B as \href{http://dx.doi.org/10.1016/j.physletb.2022.137438}{\doi{10.1016/j.physletb.2022.137438}.}}
\begin{document}\cmsNoteHeader{SMP-20-013}

\ifthenelse{\boolean{cms@external}}{\providecommand{\cmsLeft}{upper\xspace}}{\providecommand{\cmsLeft}{left\xspace}}
\ifthenelse{\boolean{cms@external}}{\providecommand{\cmsRight}{lower\xspace}}{\providecommand{\cmsRight}{right\xspace}}
\newcommand{\wg}{\ensuremath{\PW{}\Pgg}\xspace}
\newcommand{\zg}{\ensuremath{\PZ{}\Pgg}\xspace}
\newcommand{\sobs}{\ensuremath{\sigma^{\text{obs}}}\xspace}
\newcommand{\ssm}{\ensuremath{\sigma^{\text{SM}}}\xspace}
\newcommand{\mt}{\ensuremath{m_{\mathrm{T}}}\xspace}
\newcommand{\mjjvbs}{\ensuremath{m_{\mathrm{jj}}^{\mathrm{VBS}}}\xspace}
\newcommand{\detavbs}{\ensuremath{\Delta \eta_{\mathrm{jj}}^{\mathrm{VBS}}}\xspace}
\newcommand{\mtw}{\ensuremath{\mt^{\PW}}\xspace}
\newcommand{\wlpt}{\ensuremath{\pt^{\PW,\ell}}\xspace}
\newcommand{\vhad}{\ensuremath{\PV_{\mathrm{had}}}\xspace}
\newcommand{\mew}{\ensuremath{\mu_{\mathrm{EW}}}\xspace}
\newcommand{\mqcd}{\ensuremath{\mu_{\mathrm{QCD}}}\xspace}
\newcommand{\mvhad}{\ensuremath{m_{\PV}}\xspace}
\newcommand{\unsim}{\mathord{\sim}}

\cmsNoteHeader{SMP-20-013}

\title{Evidence for \texorpdfstring{$\PW\PW/\PW\PZ$}{WW/WZ} vector boson scattering in the decay channel \texorpdfstring{$\ell\nu\PQq\PQq$}{ellnuqq} produced in association with two jets in proton-proton collisions at \texorpdfstring{$\sqrt{s}=13\TeV$}{sqrt(s) = 13 TeV}}

\date{\today}

\abstract{Evidence is reported for electroweak (EW) vector boson scattering in the decay channel $\ell\nu\PQq\PQq$ of two weak vector bosons $\PW\PV$ ($\PV=\PW$ or $\PZ$), produced in association with two parton jets. The search uses a data set of proton-proton collisions at 13\TeV collected with the CMS detector during 2016--2018 with an integrated luminosity of 138\fbinv. Events are selected requiring one lepton (electron or muon), moderate missing transverse momentum, two jets with a large pseudorapidity separation and a large dijet invariant mass, and a signature consistent with the hadronic decay of a $\PW/\PZ$ boson. The cross section is computed in a fiducial phase space defined at parton level requiring all parton transverse momenta $\pt > 10\GeV$ and at least one pair of outgoing partons with invariant mass $m_{\PQq\PQq}>100\GeV$. The measured and expected EW $\PW\PV$ production cross sections are $1.90^{+0.53}_{-0.46}$\unit{pb} and $2.23^{+0.08}_{-0.11}(\text{scale}) \pm0.05(\text{PDF})$\unit{pb}, respectively, where PDF is the parton distribution function. The observed EW signal strength is $\mew = 0.85\pm0.12\,(\text{stat})^{+0.19}_{-0.17}\,(\text{syst})$, corresponding to a signal significance of 4.4 standard deviations with 5.1 expected, and it is measured keeping the quantum chromodynamics (QCD) associated diboson production fixed to the standard model prediction. This is the first evidence of vector boson scattering in the $\ell\nu\PQq\PQq$ decay channel at LHC. The simultaneous measurement of the EW and QCD associated diboson production agrees with the standard model prediction.}

\hypersetup{
pdfauthor={CMS Collaboration}, 
pdftitle={Evidence for WW/WZ vector boson scattering in the decay channel ellnuqq produced in association with two jets in proton-proton collisions at sqrt(s) = 13 TeV},
pdfsubject={CMS}, 
pdfkeywords={CMS, vector boson scattering}}

\maketitle

\section{Introduction}

The discovery of the Higgs boson~\cite{Aad:2012tfa,Chatrchyan:2012xdj} completed the observation of the
particle content of the standard model (SM) of fundamental interactions, but the investigation of its scalar and Yukawa
sectors is still in its infancy with respect to the vast scientific program foreseen with the data that is being
delivered by the Large Hadron Collider (LHC) at CERN.

Vector boson scattering (VBS) plays a special role, since the violation of its unitarity coming from direct interaction
between vector bosons is prevented by counterbalancing diagrams involving the Higgs boson~\cite{Lee:1977eg}.  This
precise cancellation of divergent effects is an important aspect of the SM, and one of the main motivations to study the
VBS processes.

In fact, the VBS measurements could provide an additional insight into the electroweak (EW) symmetry breaking, as well as
a powerful tool to test effects beyond the SM that can perturb the delicate equilibrium present in the total cross
section calculation.  The VBS production of vector boson pairs is rare at the LHC, since it is a purely EW process of
order 6 of the neutral weak current coupling $\alpha_{\mathrm{EW}}^6$, and it has a large background contamination.
Only in recent years the data set collected by the LHC experiments has become large enough to permit measurements in
fully leptonic final states~\cite{Sirunyan:2017ret, Aaboud:2019nmv,Sirunyan:2019ksz, Aaboud:2018ddq, ATLAS:2020nlt,
Sirunyan:2017fvv} and in the $\PZ\PGg$ channel~\cite{CMS:2021gme,ATLAS:2019qhm} by the ATLAS~\cite{ATLAS:2008xda} and
CMS~\cite{Chatrchyan:2008zzk} Collaborations.  At the same time, the theory community showed a renewed interest in the
vector boson scattering~\cite{Alessandro:2018khj}, both in terms of the SM measurements and searches for beyond SM
effects.  Therefore, it is compelling to study all the VBS final states accessible at the LHC in addition to the fully
leptonic ones and those with photons.

\begin{figure*}[htb]
  \centering
  \includegraphics[width=0.33\textwidth]{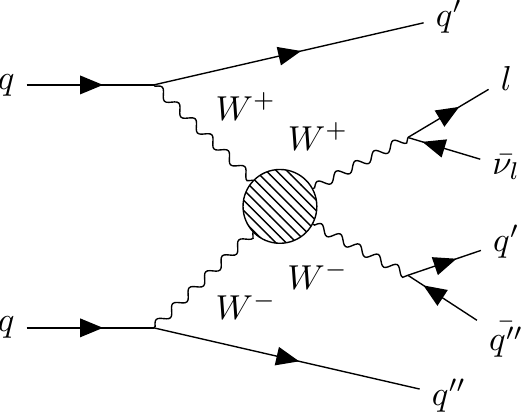}
  \hspace{2cm}
  \includegraphics[width=0.33\textwidth]{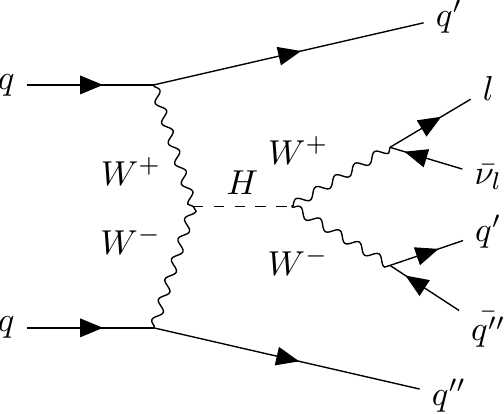}
  \vspace{1cm}
  \includegraphics[width=0.33\textwidth]{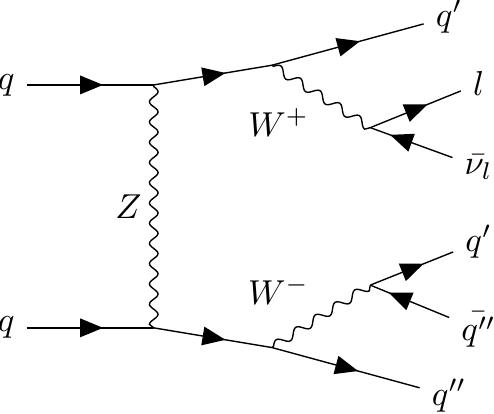}
  \hspace{2cm}
  \includegraphics[width=0.33\textwidth]{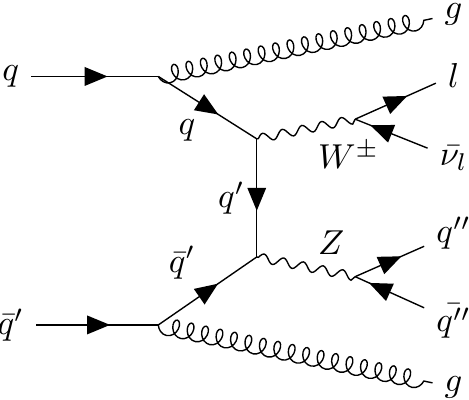}

\caption{Examples of Feynman diagrams contributing to the analyzed final state:
general schema of purely EW VBS signal process contributions (upper left diagram), the $s$-channel Higgs boson
contribution (upper right diagram), the purely EW nonresonant diboson production (lower left diagram); and an example of
non-EW nonresonant diboson production (lower right diagram), which is part of the irreducible
background. } \label{fig:diagram}
\end{figure*}

In this letter, we address the case where one of the vector bosons decays into quarks, whereas the other one, a {\PW}
boson, decays into a lepton $\ell$ (electron or a muon), and a neutrino.  Fig.~\ref{fig:diagram} shows examples of the
Feynman diagrams describing some processes contributing to this final state.

Both ATLAS and CMS Collaborations have already studied the VBS $\PW\PV$ process, where {\PV} stands for a {\PW} or {\PZ}
boson, in final states where one boson decays leptonically, and the other boson decays hadronically, with data collected
in 2016, corresponding to only one quarter of the full Run~2 dataset obtained from 2016 to 2018.  The CMS
analysis~\cite{Sirunyan:2019der} was focused on a search for anomalous EW $\PW\PV$ production and put stringent limits
on effective field theory operators, whereas the ATLAS paper~\cite{Aad:2019xxo} reached a SM signal significance of 2.7
standard deviations including the $\ell\ell\PQq\PQq$ and $\nu\nu\PQq\PQq$ decay channels. ATLAS also studied
anomalous couplings in the $\PW\PW/\PW\PZ jj$ channel~\cite{ATLAS:2016nmw} using 8 TeV data from the 2012 dataset.
This paper reports the first evidence for the SM EW $\ell\nu\PQq\PQq$
plus two jets production by analyzing the full LHC
data set collected by CMS between 2016 and 2018, corresponding to an
integrated luminosity of $138\fbinv$.

\section{Signal and background simulation}\label{sec:2}

The signal is characterized by the presence of a single isolated electron or muon, a moderate amount of missing
transverse momentum \ptmiss, and either three or four jets.  One pair of jets is required to have a large invariant mass
and large pseudorapidity ($\eta$) separation, the typical signature of VBS-like events, whereas the remaining jets are
the result of a vector boson decay.  If the boson has a high enough momentum in the laboratory reference frame, its
decay products can be collected in a single jet, whereas at lower momentum the decay is resolved into two separate jets.

The main sources of background contamination originate from the production of a single {\PW} boson accompanied by jets
(called {\PW}+jets in the following), and \ttbar pairs, where one of the {\PW} bosons produced by the top quark decays
hadronically.  Although simulated samples for these backgrounds are available, an approach based on control samples from
data is applied to improve the description of these backgrounds in the signal region.  The {\PW}+jets contribution from
Monte Carlo (MC) is corrected differentially by exploiting the events in a dedicated control region, which is described
in more detail in Section~\ref{sec:5}. The kinematic distributions of the top quark background are taken
from MC, but the normalization is measured from data in the dedicated control region.

The following background processes are modeled using MC event generators: nonresonant QCD-associated diboson production
(QCD-$\PW\PV$);
\ttbar and single top quark production (in $s$, $t$ channel and $\PQt\PW$);
Drell--Yan (DY) lepton pair production; {\PV} boson production in association with a photon (\wg and \zg); single vector
boson EW production in the vector boson fusion channel (VBF-{\PV}); and triboson production ($\PV\PV\PV$).  The QCD
multijet background, which may enter the signal region with nonprompt leptons, is estimated instead in a purely
data-driven method, as described in Sec.~\ref{sec:5}.

Most of the processes are simulated at next-to-leading order (NLO) in the strong coupling $\alpha_{s}$ using
\POWHEG~v2~\cite{Frixione:2002ik,Nason:2004rx,Frixione:2007vw,Alioli:2008gx,Alioli:2010xd}, 
\MGvATNLO v2.4.2 ~\cite{Alwall:2014hca,Alwall:2007fs}, 
or \MCFM~v7.0~\cite{campbell_precision_2019,campbell_multi-threaded_2015,campbell_vector_2011,campbell_update_1999}. 
Only the {\PW}+jets, QCD-$\PV\PV$, VBF-{\PV}, and \wg events are generated with \MGvATNLO v2.4.2 at leading order (LO)
accuracy in perturbative quantum chromodynamics (QCD).  The \ttbar component of the top quark background and the DY
events are also weighted using generator level information to improve the agreement of the simulated transverse momentum
(\pt) distributions of the \ttbar and DY systems~\cite{Czakon:2017wor,CMS:2016oae,CMS:2019raw} to data.

The signal, namely the VBS $\PW(\ell\nu)\PV(\mathrm{jj})$ process (the parentheses give the decay modes), is simulated
with \MGvATNLO v2.6.5 at leading order: the intermediate-state vector boson pair is produced by implementing the narrow
width approximation and then decayed by MadSpin~\cite{Madspin} to partially account for finite-width effects and spin
correlations.  The contribution from the VBS $\PZ(\ell\ell)\PV(\mathrm{jj})$ production, where one of the leptons falls
beyond the acceptance of the analysis, is considered as a background.  The {\PW} decay into \PGt is considered part of
the signal, and {\PW} decays into all leptonic final states are generated. However, the analysis has been performed
only for electron and muon final states.

Apart from VBF-{\PV}, all MC samples for the parton showering, hadronization, and the simulation of the underlying event
are provided by \PYTHIA~8.226 (8.230) \cite{Sjostrand:2014zea,Alioli:2010xd}.  The NNPDF 3.0 NLO \cite{Ball:2014uwa}
(NNPDF 3.1 NNLO \cite{Ball:2017nwa}) parton distribution functions (PDFs) are used for simulating all 2016 (2017 and
2018) samples.  The modeling of the underlying event is generated using the CUETP8M1 \cite{Skands:2014pea,CMS:2015wcf}
(CP5 \cite{CMS:2019csb}) tune for simulated samples corresponding to the 2016 (2017 and 2018) data.

To improve the description of the additional jet emissions by the parton shower simulation in the VBS
topology~\cite{covarelli_vector-boson_2021,ballestrero_precise_2018,Jager:2020hkz}, the dipole recoil scheme is used in
 \PYTHIA for the VBS signal MC sample, and the \HERWIG~7.0~\cite{Bahr:2008pv,Bellm:2015jjp}
program is used for the VBF-{\PV} background.

The interference between the EW and QCD diagrams for the $\PW^{\pm}\PW^{\pm}$, $\PW^{\pm}\PW^{\mp}$, and $\PW\PZ$
 processes, generated with $\MGvATNLO$ including the contributions of order $\alpS\alpha_{\mathrm{EM}}^5$, is less than
 3\% of the signal inclusively in the phase space region of interest of the analysis and is neglected.
 
For all processes, the detector response is simulated using a detailed description of the CMS detector, based on
the \GEANTfour package~\cite{Agostinelli:2002hh}.  Additional interactions in the same or adjacent bunch crossings
(pileup) based on minimum bias events simulated with the \PYTHIA
shower simulation are overlaid onto each event, with the number of
interactions drawn from a distribution that is similar to the one observed in data.  The average number of such
interactions per event is $\approx$23 and $\approx$32 for the 2016 data and 2017-2018 data, respectively.

\section{The CMS detector}\label{sec:3}

A detailed description of the CMS detector, together with a definition of the coordinate system used and the relevant
kinematic variables, can be found in Ref.~\cite{Chatrchyan:2008zzk}.

The central feature of the CMS apparatus is a superconducting solenoid of 6\unit{m} internal diameter, providing a
magnetic field of 3.8\unit{T}. Within the solenoid volume are a silicon pixel and strip tracker, a lead tungstate
crystal electromagnetic calorimeter (ECAL), and a brass and scintillator hadron calorimeter (HCAL), each composed of a
barrel and two endcap sections. Forward calorimeters extend the pseudorapidity coverage provided by the barrel and
endcap detectors. Muons are measured in gaseous detectors embedded in the steel flux-return yoke outside the solenoid.
The particle-flow algorithm~\cite{CMS-PRF-14-001} reconstructs and identifies each individual particle in an event, with
an optimized combination of information from the various elements of the CMS detector.  The energy of photons is
obtained from the ECAL measurement.  The energy of electrons is determined from a combination of the electron momentum
at the primary interaction vertex as measured by the tracker, the energy of the corresponding ECAL cluster, and the
energy sum of all bremsstrahlung photons spatially compatible with originating from the electron track.  The energy of
muons is obtained from the curvature of the corresponding track.  The energy of charged hadrons is determined from a
combination of their momentum measured in the tracker and the matching ECAL and HCAL energy deposits, corrected for the
response function of the calorimeters to hadronic showers.  Finally, the energy of neutral hadrons is obtained from the
corresponding corrected ECAL and HCAL energies.

Events of interest are selected using a two-tiered trigger system.  The first level, composed of custom hardware
processors, uses information from the calorimeters and muon detectors to select events at a rate of up to 100\unit{kHz}
within a fixed latency of about 4\mus~\cite{Sirunyan:2020zal}.  The second level, known as the high-level trigger,
consists of a farm of processors running a version of the full event reconstruction software optimized for fast
processing, and reduces the event rate to around 1\unit{kHz} before data storage~\cite{Khachatryan:2016bia}.

\section{Event reconstruction, selection and categorization}\label{sec:4}

Events are selected for further analysis by triggers for isolated single leptons with \pt thresholds of 27, 32, 35\GeV
for electrons and of 24, 24, 27\GeV for muons, respectively for the 2016, 2017, 2018 data-taking periods.  The final
leptons are required to have an offline reconstructed \pt of at least $35\GeV$ ($30\GeV$) for electron (muon)
candidates, and a pseudorapidity of $\abs{\eta}<2.5$ (2.4) for electrons (muons).

For each event, hadronic jets are clustered from reconstructed particles using the infrared- and collinear-safe anti-\kt
algorithm~\cite{Cacciari:2008gp, Cacciari:2011ma} with a distance parameter of 0.4 (0.8), labeled in the following as
AK4 (AK8) jets.  Additional proton-proton interactions within the same or nearby bunch crossings (pileup) can contribute
additional tracks and calorimetric energy depositions to the jet momentum.  To mitigate this effect, charged particles
identified as originating from pileup vertexes are discarded and an offset correction is applied to correct for
remaining contributions. Reconstructed jets cannot overlap with isolated leptons:
$\Delta R(j, \ell)= \sqrt{\smash[b]{(\Delta\eta)^2 + (\Delta\phi)^2}} >$ 0.4 (0.8) for AK4 (AK8) jets. 

In an event, AK4 and AK8 jets are considered in the analysis if they have a $\pt>30\GeV$ and $\abs{\eta}<4.7$ or
$\pt>200\GeV$ and $\abs{\eta}<2.4$, respectively.  The pileup-per-particle identification algorithm
(PUPPI)~\cite{Sirunyan:2020foa,Bertolini:2014bba} is applied to AK8 jet constituents to remove pileup tracks at the
reconstructed particle level.  Moreover, a grooming algorithm, known as ``soft drop''
(SD)\cite{Dasgupta:2013ihk,Butterworth:2008iy,Larkoski:2014wba}, is applied to the constituents of AK8 jets reclustered
using the Cambridge--Aachen algorithm~\cite{Dokshitzer:1997in,Wobisch:1998wt}.  The SD algorithm, which has as an
angular exponent $\beta = 0$, soft cutoff threshold $z_{\text{cut}} < 0.1$, and characteristic radius $R_{0} =
0.8$~\cite{Larkoski:2014wba}, removes soft, wide-angle radiation from the large radius jet, improving the modeling of
the jet mass observable.  The parameters of the SD algorithm are calibrated in a top quark-antiquark sample enriched in
hadronically decaying $\PW$ bosons~\cite{Khachatryan:2014vla}.  The AK8 jets are identified as hadronic decays of
Lorentz-boosted $\PW/\PZ$ bosons using the ratio between 2- and 1-subjettiness~\cite{Thaler:2010tr} variables denoted as
$\tau_{21}=\tau_{2}/\tau_{1} < 0.45$ and a groomed AK8 jet mass between 40 and 250\GeV. AK4 jets overlapping
with AK8 jets that pass the preselection with $\Delta R(j_{\text{AK4}}, j_{\text{AK8}}) < 0.8$ are removed from the event.  

The analysis targets the VBS production of pairs of vector bosons, $\PW\PV$, in association with two jets originating
from the scattered incoming partons, called tag jets.  In the chosen signal process, the {\PW} boson decays leptonically
and the second boson decays hadronically.  Candidate events are required to contain exactly one tightly identified and
isolated lepton \cite{Khachatryan:2015hwa,Chatrchyan:2012xi} associated with the {\PW} boson leptonic decay. Events
containing a second loosely identified lepton with $\pt > 10\GeV$ are vetoed.  Finally, we require a missing transverse
momentum $\ptmiss > 30\GeV$ in the event.  The missing transverse momentum vector \ptvecmiss is computed as the negative
vector sum of the transverse momenta of all the particle candidates in an event~\cite{Sirunyan:2019kia}.  The PUPPI
algorithm is also applied to reduce the pileup dependence of the \ptmiss observable. The \ptmiss vector is computed from
the particle-flow candidates weighted by their probability to originate from the primary interaction
vertex~\cite{Sirunyan:2019kia}.

Two main categories are defined depending on the reconstruction regime of the hadronically decaying vector boson.  An
event is assigned to a boosted category if it contains only one AK8 jet, with $\pt > 200\GeV$ and $|\eta|<2.4$, that
passes the selection criteria as a hadronically decaying vector boson \vhad, together with at least two AK4
jets. Otherwise, if no AK8 jet V boson candidate is found and instead at least four AK4 jets are reconstructed with $\pt
> 30\GeV$, the event is assigned to a resolved category.  In both resolved and boosted categories, the two AK4 jets with
the largest invariant mass are identified as the VBS tag jets.  In the resolved category, out of the remaining jets
after the VBS tag jet selection, the two jets with invariant mass closest to $85\GeV$ (the average between the {\PW} and
{\PZ} boson masses) are chosen as the decay product of \vhad.

The fraction of VBS events in the sample is enhanced requiring a large invariant mass $\mjjvbs > 500\GeV$ and large
pseudorapidity interval $\detavbs= | \eta_{j1}^{\text{VBS}} - \eta_{j2}^{\text{VBS}}| > 2.5$ for the tag jets.  The
leading VBS tag jet is required to have $\pt > 50\GeV$ and the transverse mass of the leptonically decaying {\PW} is
required to be $\mtw < 185\GeV$, defined as
\begin{linenomath}
\begin{equation}
\mtw = \sqrt{2\,\pt(\ell) \ptmiss [ 1 - \cos (\Delta\varphi (\pt(\vec{\ell}), \ptvecmiss))]}~,
\end{equation}
\end{linenomath}
where $\pt(\ell)$ is the \pt of the lepton and $\Delta\varphi (\pt(\vec{\ell}), \ptvecmiss)$ is the azimuthal distance
between the lepton and the \ptvecmiss.

After these selections, the signal and control regions for the main backgrounds, the top quark and {\PW}+jets ones, are
defined in both the resolved and boosted regions in a similar manner.

The signal region consists of events where:
(i) no {\PQb} jet candidates are found according to the loose working point
of the \textsc{DeepCSV} tagger~\cite{Sirunyan:2017ezt},
(ii) a machine-learning {\PQb}-tagging algorithm with a
{\PQb}-tagging efficiency $\ge85\%$ and a mistag probability $\le20\%$ reduces the contamination
from \ttbar events,
and (iii) the hadronically decaying vector boson
invariant mass $\mvhad$ is between $65-105$ ($70-115$)\GeV for the resolved (boosted) category, which is consistent with
an on-shell {\PW} or {\PZ} decaying hadronically.  Events falling in the same $\mvhad$ interval as the signal but
containing at least one {\PQb} jet are classified in the top quark control region.  Finally, if no {\PQb} jets exist and
$\mvhad$ is not within the $\PW$ or $\PZ$ resonance range, $\mvhad\notin(65 , 105)\GeV$ or $\mvhad\notin(70 , 115)\GeV$
for the resolved and boosted cases, events are classified as part of the {\PW}+jets control region.  All of the signal
and control regions are split according to the flavor of the selected lepton (electron or muon).

The purity of the top quark processes in the dedicated control regions is $\unsim 90\%$, where the largest
contaminations are due to the nonprompt ($\unsim6\%$) and W+jets ($\unsim3\%$) backgrounds.  The purity of the W+jets
process in the dedicated control regions is between $\unsim50\%$ in the resolved categories and $\unsim60\%$ in the
boosted ones. The largest contaminations are due to the nonprompt background ($\unsim28\%$ in the resolved and
$\unsim15\%$ in the boosted category), DY processes ($\unsim10\%$), and top quark production ($\unsim6\%$).

\section{Background estimation}\label{sec:5}

The largest background contribution is the {\PW}+jets process, followed by the top quark and the QCD multijet
backgrounds. The contamination from the single vector boson EW production in the VBF channel (VBF-{\PV}) is negligible
in the resolved category (2\% inclusively), but more important in the boosted one (4\% inclusively).

The {\PW}+jets contribution is corrected using control samples in data.  It is experimentally observed that the
transverse momentum of the leptonically decaying {\PW} boson (\wlpt), measured using the lepton momentum and
the \ptmiss, and the VBS tag jets \pt are poorly described by simulation in the multijet phase space region used in this
analysis.  To correct this important background in a differential way, the {\PW}+jets MC sample is split into several
subsamples according to \wlpt (both categories) and \pt of the subleading VBS tag jet (only in the resolved category),
and their normalizations are left floating and uncorrelated in the final fit. The data in the {\PW}+jets control region is used
to constrain the normalization of the {\PW}+jets subcomponents by including in the fit the same variables
with the same binning used to define the MC sample splitting.
As a result of this procedure, the normalization corrections to the W+jets MC
sample are propagated in the fit and also to the signal region including the full covariance matrix.

The closure test for this correction is performed by dividing the {\PW}+jets control region into two subregions, defined
by two intervals of \mvhad closer to (\ie, $[50,65] \cup [105,150] \GeV$ in the resolved category or $[50,70] \cup
[115,150] \GeV$ in the boosted one) or farther (\ie, $[40, 50] \cup [150,+\infty] \GeV$) from the {\PV} resonance where
the signal region is located.  Correction factors are derived for each {\PW}+jets component in the two subregions and
are in agreement with each other. Therefore, including the {\PW}+jets control region in the final fit to extract the
{\PW}+jets correction factors provides a meaningful description for the {\PW}+jets MC also in the signal region.

The top quark background contribution is determined from MC simulation except for its normalization, which is measured
in the top quark enriched control region in the final fit to the data.

The QCD multijet background, also called nonprompt, is estimated from data by measuring the probability for a loosely
defined reconstructed lepton originating from a jet to be misidentified as a tightly reconstructed lepton in a phase
space region outside the analysis region. The QCD-enriched region is defined by the presence of at least one lepton with
the same \pt requirement as for the rest of the analysis, $\ptmiss < 20\GeV$, $\mtw < 20\GeV$, at least one AK4 jet in
the event with $\Delta R>1$ from the lepton.  The contribution
from EW processes with a real lepton is subtracted from this QCD enriched phase space region by means of {\PW}+jets and
DY MC events.
 
The contributions from the other backgrounds, \eg, QCD-$\PV\PV$, DY, VBF-{\PV}, $\PV\PV\PV$, $\PV\gamma$,
VBS-$\PZ(\ell\ell)\PV(\mathrm{jj})$ processes, are estimated from MC simulation.

\section{Signal extraction}\label{sec:6}

\begin{figure}[!htb]
  \centering
  \includegraphics[width=0.49\textwidth]{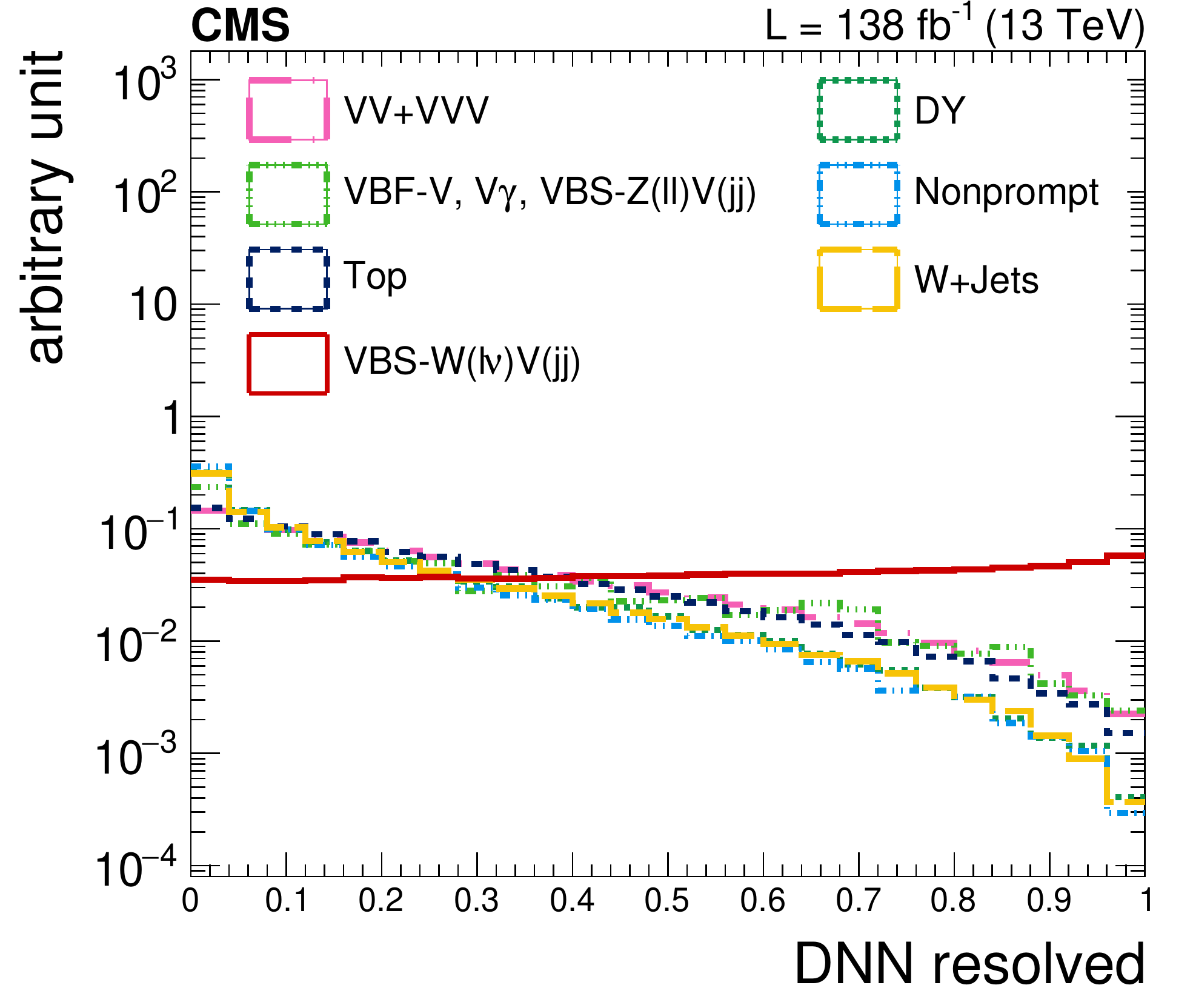}
  \includegraphics[width=0.49\textwidth]{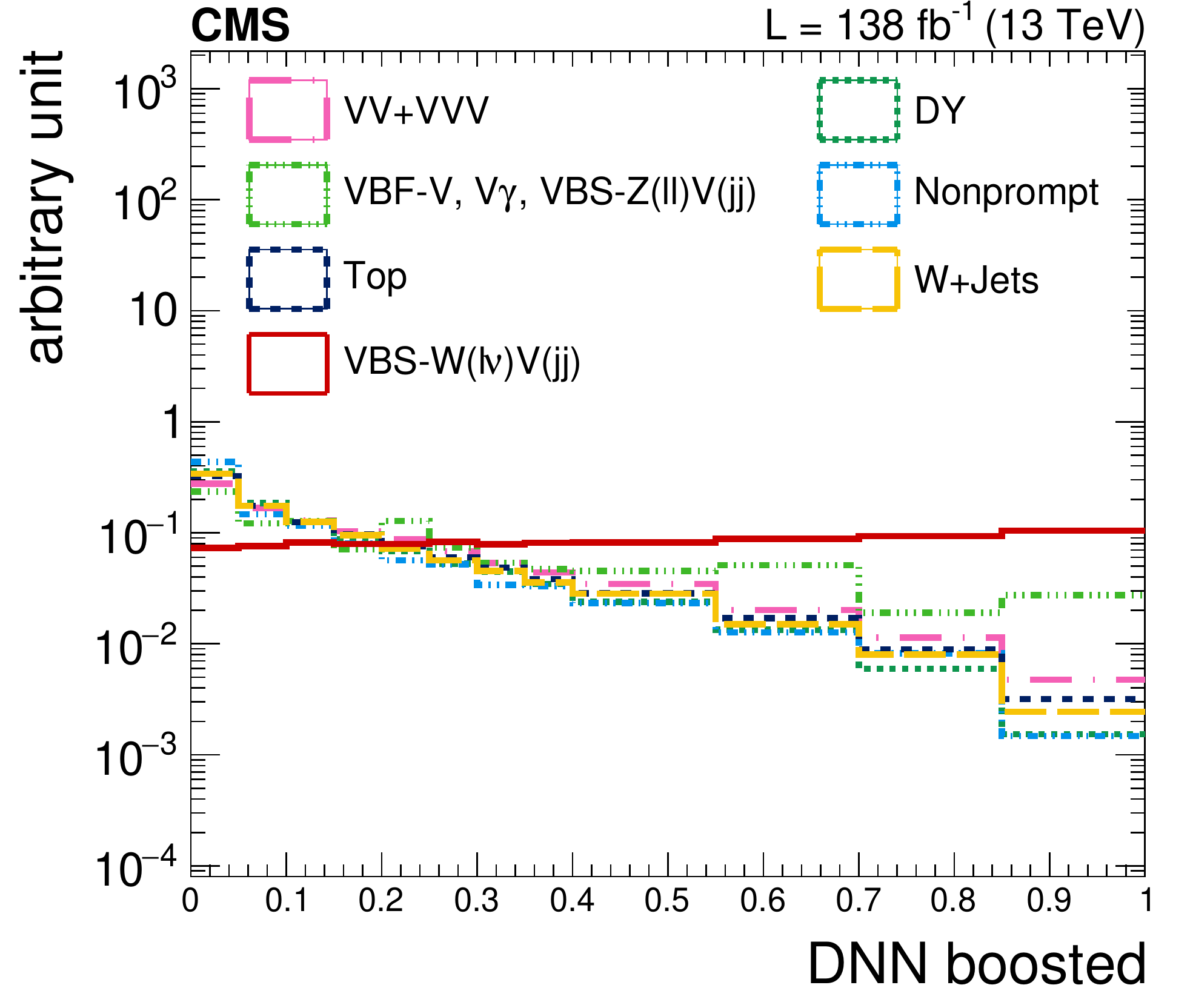}
  \caption{
    The DNN discriminator distribution, taken from simulation, for VBS signal and backgrounds in the resolved (\cmsLeft)
    and boosted (\cmsRight) signal regions normalized to unity.  } \label{fig:norm:dnnbin}
\end{figure}

Because of the large background and complex signal topology, the most significant features to separate signal and
backgrounds are condensed in a single discriminator built with a deep neural network (DNN).  This approach increases the
sensitivity of the analysis by a factor of three over a fit to the shape of the most sensitive variable \mjjvbs.  Two
different discriminators are optimized for the resolved and boosted categories since the event topology and the
kinematics change significantly between the two.  The DNN implementation consists of a fully connected neural network
with four layers with 64 (32) nodes for the resolved (boosted) topology, trained with stochastic gradient descent
implemented via the ``Adam'' optimizer \cite{kingma_adam:_2017}.  The models are trained minimizing the binary
cross-entropy~\cite{binary_crossentropy,Goodfellow-et-al-2016} loss until full convergence.  All the backgrounds are
included as a single class against the signal in the optimization of the DNN discriminator, weighted by their relative
importance. To ensure that the discrimination is not degraded by treating the backgrounds as a single class, the DNN is evaluated
separately for each type of background, and the shapes of the distributions are similar for all
backgrounds, as shown in Fig.~\ref{fig:norm:dnnbin}.
Overfitting is carefully avoided by the use of regularization techniques, such as
Dropout and L2 weights decay \cite{Goodfellow-et-al-2016}.  A technique called SHAP (SHapley Additive
exPlanations) \cite{shap,Shapley:1953ctg}, developed in the field of explainable machine learning, is applied to
cross-check the dependence of the DNN model on the input variables and to rank their importance.  Among the most
important ones, as identified by SHAP and matching the physics expectation, are the
\mjjvbs variable, the Zeppenfeld variable \cite{Rainwater:1996ud} of the lepton, and the quark/gluon discriminator 
variable of the leading \vhad jet, which is based on a likelihood discriminant constructed with three variables for
each jet: the jet energy, the multiplicity of the jet constituents, 
and the minor axis width of the ellipse in the $\eta-\phi$ plane containing the jet constituents~\cite{CMS:2013kfa,CMS-DP-2016-070}.

\begin{figure}[!htpb]
  \centering
  \includegraphics[width=0.49\textwidth]{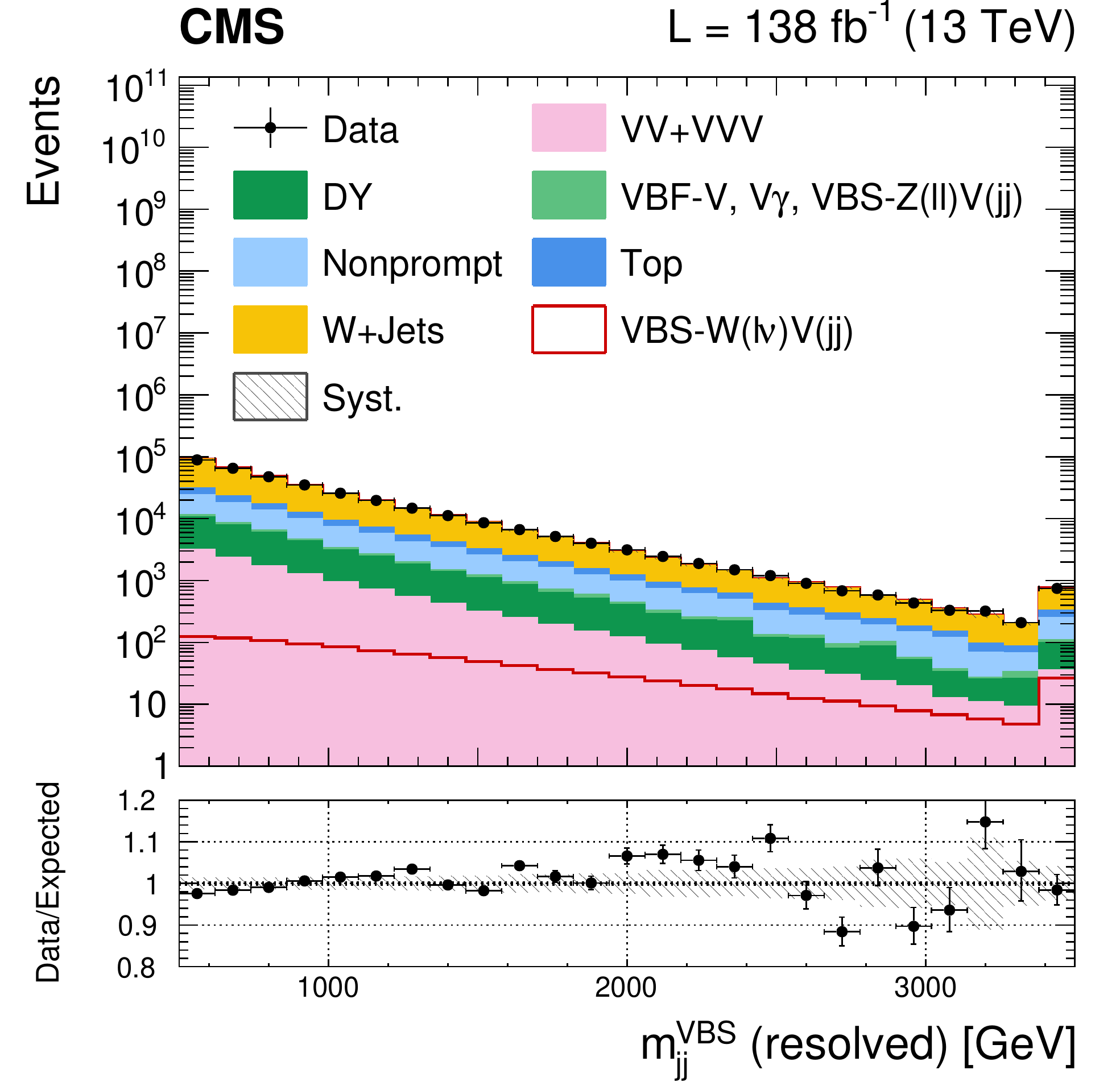}
  \includegraphics[width=0.49\textwidth]{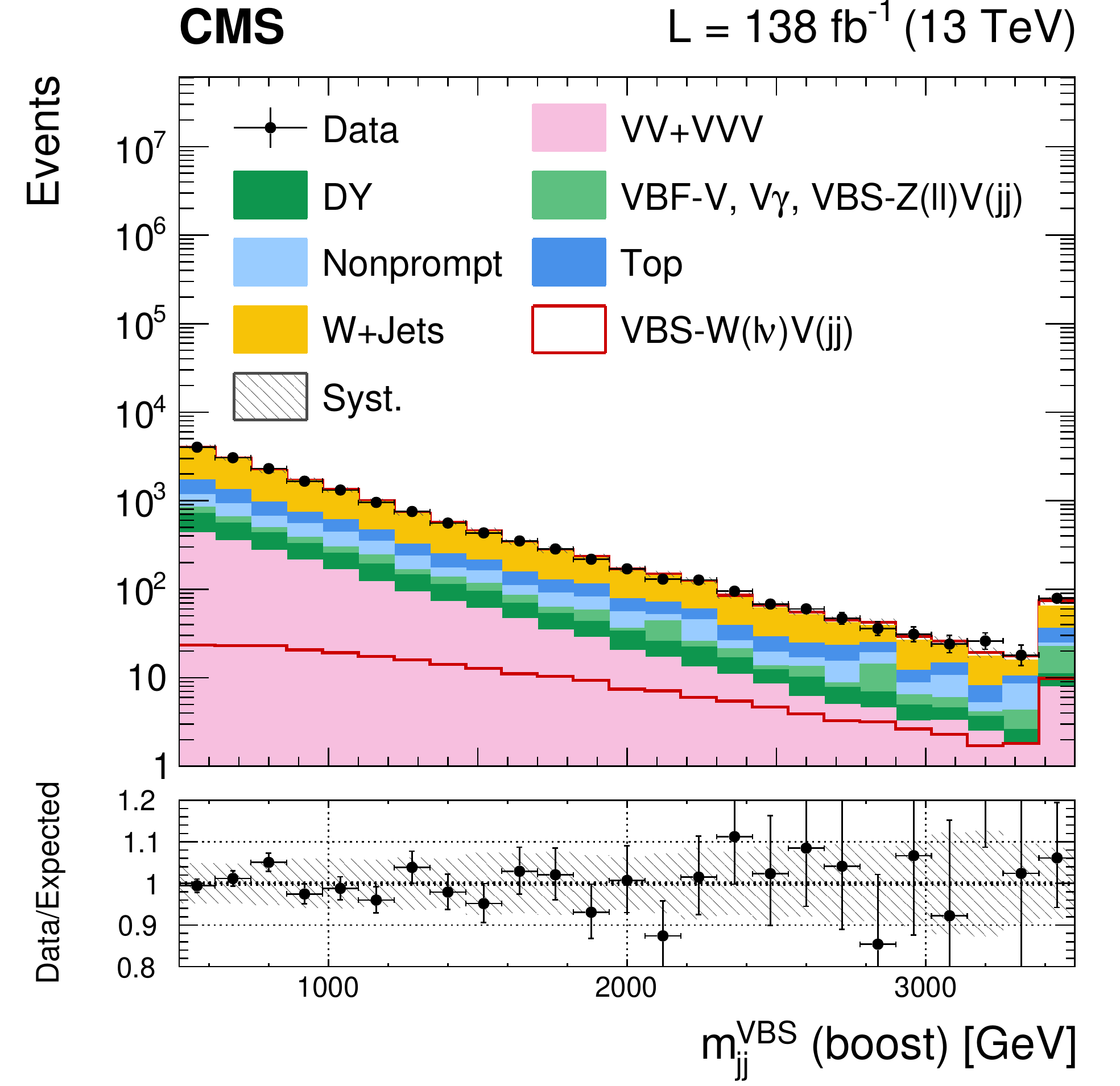}
  \caption{
    Post-fit distributions of the \mjjvbs observable in the resolved (\cmsLeft) and boosted (\cmsRight) signal regions.
    Vertical bars on data points show the statistical error, whereas the gray band is the post-fit uncertainty on MC
    with all systematic uncertainties included.  } \label{fig:mjjvbs}
\end{figure}

The post-fit distribution of the \mjjvbs variable is shown in Fig.~\ref{fig:mjjvbs}. The effect on the DNN distribution
of the small residual mismodeling observed in the post-fit \mjjvbs distribution in the resolved category, was
investigated and is negligible. In fact, the signal extraction is focused on the high DNN region, which is
strongly correlated with the high \mjjvbs range. The residual discrepancy ($<5\%$) at low \mjjvbs values does not bias the signal extraction performance.

Table~\ref{tab:variablesDefinition} shows the complete list of input variables used for the resolved and boosted
topologies, along with their ranking from the SHAP algorithm.  The Zeppenfeld variable of a particle X is defined as:
\begin{linenomath}
\begin{equation}
  \mathrm{Z}_{\mathrm{X}} = \frac{\eta^{\mathrm{X}} - \bar{\eta}^{\mathrm{VBS}}}{\detavbs},
\end{equation}
\end{linenomath}
where $\bar{\eta}^{\mathrm{VBS}}$ is the mean pseudorapidity of the VBS tag jets.

The centrality~\cite{atlas_collaboration_measurement_2017,Aaboud:2018ddq} variable is defined as $C_{\PV\PW}
= \min(\Delta\eta_- , \Delta\eta_+)$, with $\Delta\eta_{+} = \max(\eta^{\mathrm{VBS}}) - \max(\eta^{\vhad}, \eta^{\PW})$
and $\Delta\eta_{-} = \min(\eta^{\mathrm{VBS}}) - \min(\eta^{\vhad}, \eta^{\PW})$.  The $\eta^{\PW}$ value is determined
assuming the {\PW} boson mass from the lepton and \ptmiss kinematics.

\begin{table*}[!htb]
\centering
  \topcaption{ Variables used as input to the DNN for the resolved and boosted models.  They are ranked by their
    contributions to the signal discrimination power of the DNN model using the SHAP \cite{shap,Shapley:1953ctg}
    technique and their rank is shown in the table for the resolved and boosted categories
    models.  \label{tab:variablesDefinition} }
\begin{tabular}{l  c c  c c }
\multirow{2}{*}{Variable}        & \multirow{2}{*}{Resolved}   & \multirow{2}{*}{Boosted}  & \multicolumn{2}{c}{SHAP ranking}  \\
 & & & Resolved & Boosted \\
\hline
Lepton pseudorapidity & \checkmark & \checkmark & 13 & 12 \\ Lepton transverse momentum & \checkmark & \checkmark & 16 &
10 \\ Zeppenfeld variable for the lepton & \checkmark & \checkmark & 2 & 2 \\ Number of jets with $\pt > 30\GeV$
& \checkmark & \checkmark & 7 & 3 \\ Leading VBS tag jet \pt & - & \checkmark & - & 11 \\ Trailing VBS tag jet \pt
& \checkmark & \checkmark & 7 & 6 \\ Pseudorapidity interval \detavbs between tag jets & \checkmark & \checkmark & 4 &
4 \\ Quark/gluon discriminator of leading VBS tag jet& \checkmark & \checkmark & 9 & 7 \\ Azimuthal angle distance
between VBS tag jets & \checkmark & - & 10 & - \\ Invariant mass of the VBS tag jets pair & \checkmark & \checkmark & 1
& 1 \\
\pt of the leading \vhad jet  & \checkmark & - & 14 & - \\
\pt of the trailing \vhad jet  & \checkmark & - & 12 & - \\
Pseudorapidity difference between \vhad jets & \checkmark & - & 8 & - \\ Quark/gluon discriminator of the leading \vhad
jet & \checkmark & - & 3 & - \\ Quark/gluon discriminator of the trailing \vhad jet & \checkmark & - & 5 & - \\
\pt of the AK8 \vhad jet candidate   &   -  &  \checkmark   &  - & 8  \\
Invariant mass of \vhad & \checkmark & \checkmark & 11 & 5 \\ Zeppenfeld variable for \vhad & - & \checkmark & - & 9 \\
Centrality & - & \checkmark & 15 & 13 \\
\hline
\end{tabular}
\end{table*}

\begin{figure*}[!htb]
  \centering
  \includegraphics[width=0.49\textwidth]{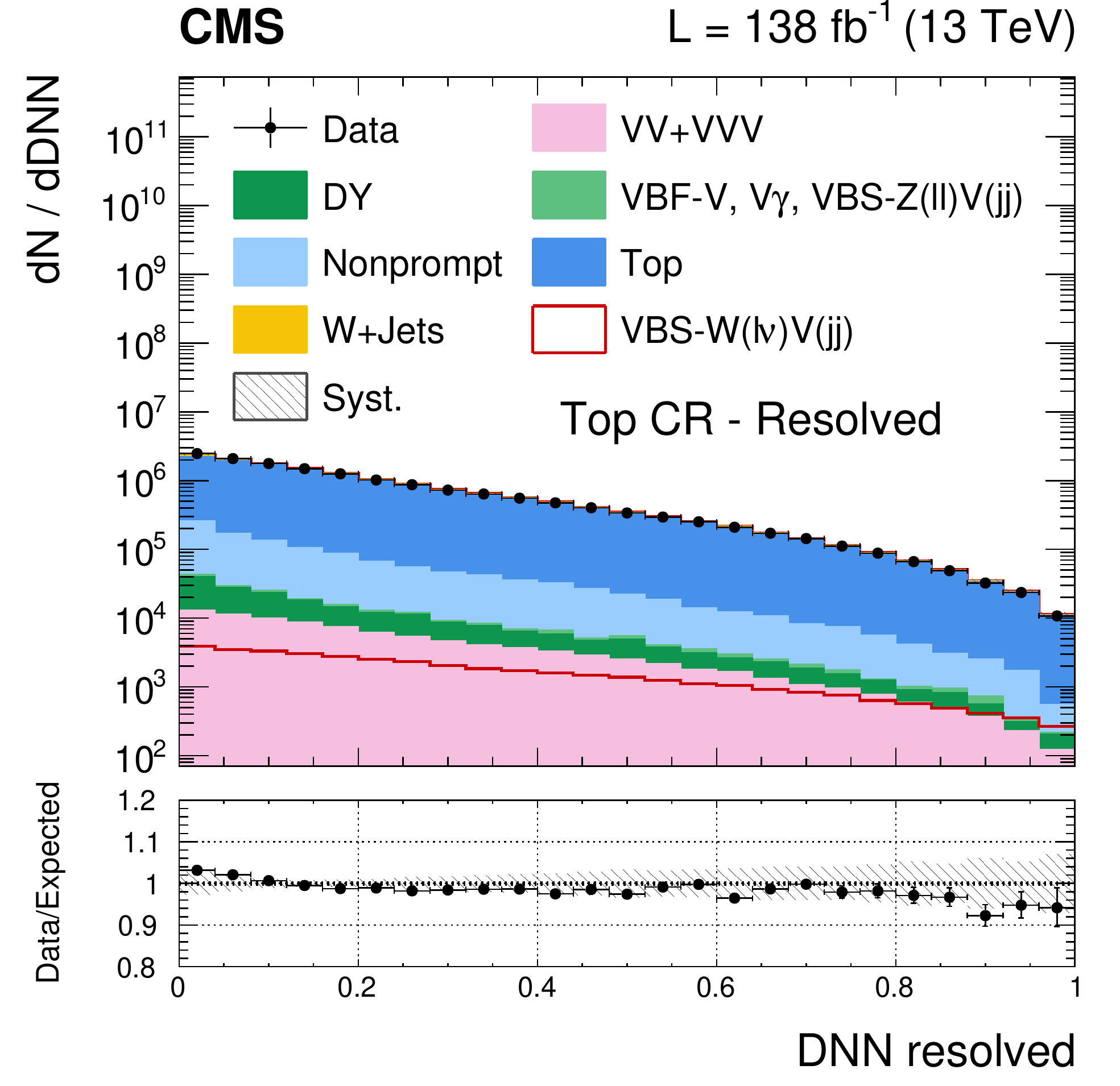}
  \includegraphics[width=0.49\textwidth]{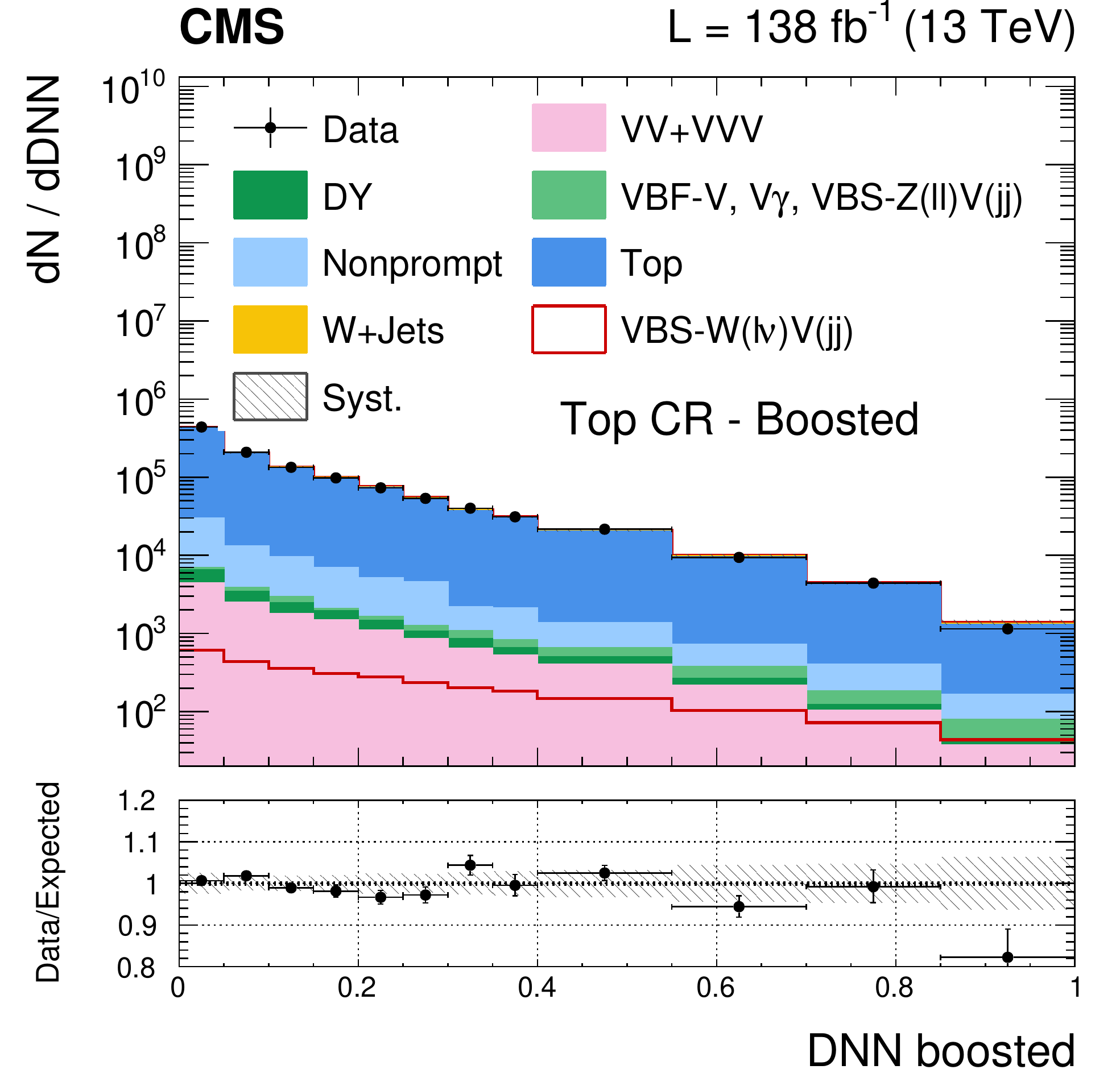}
  \includegraphics[width=0.49\textwidth]{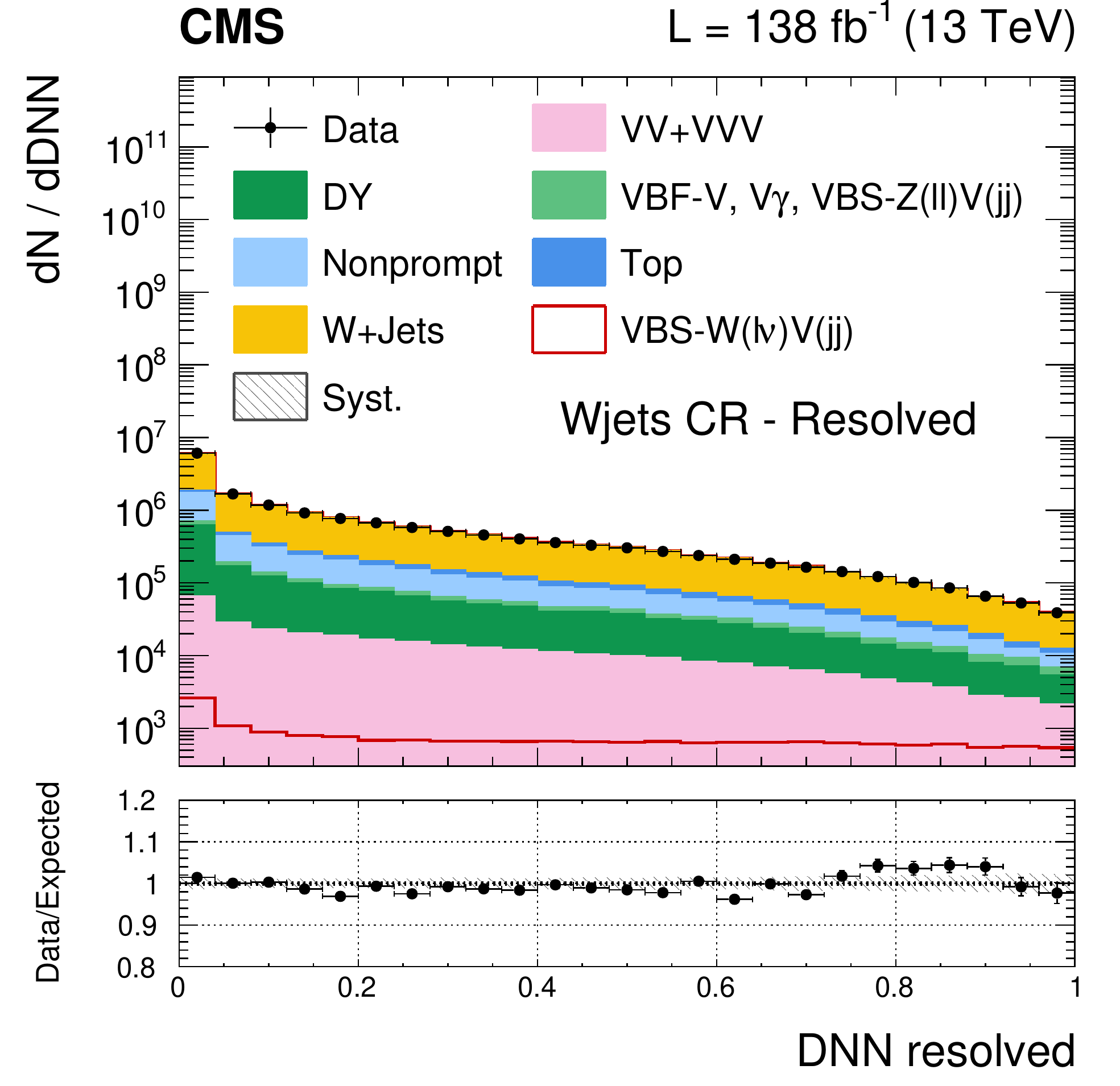}
  \includegraphics[width=0.49\textwidth]{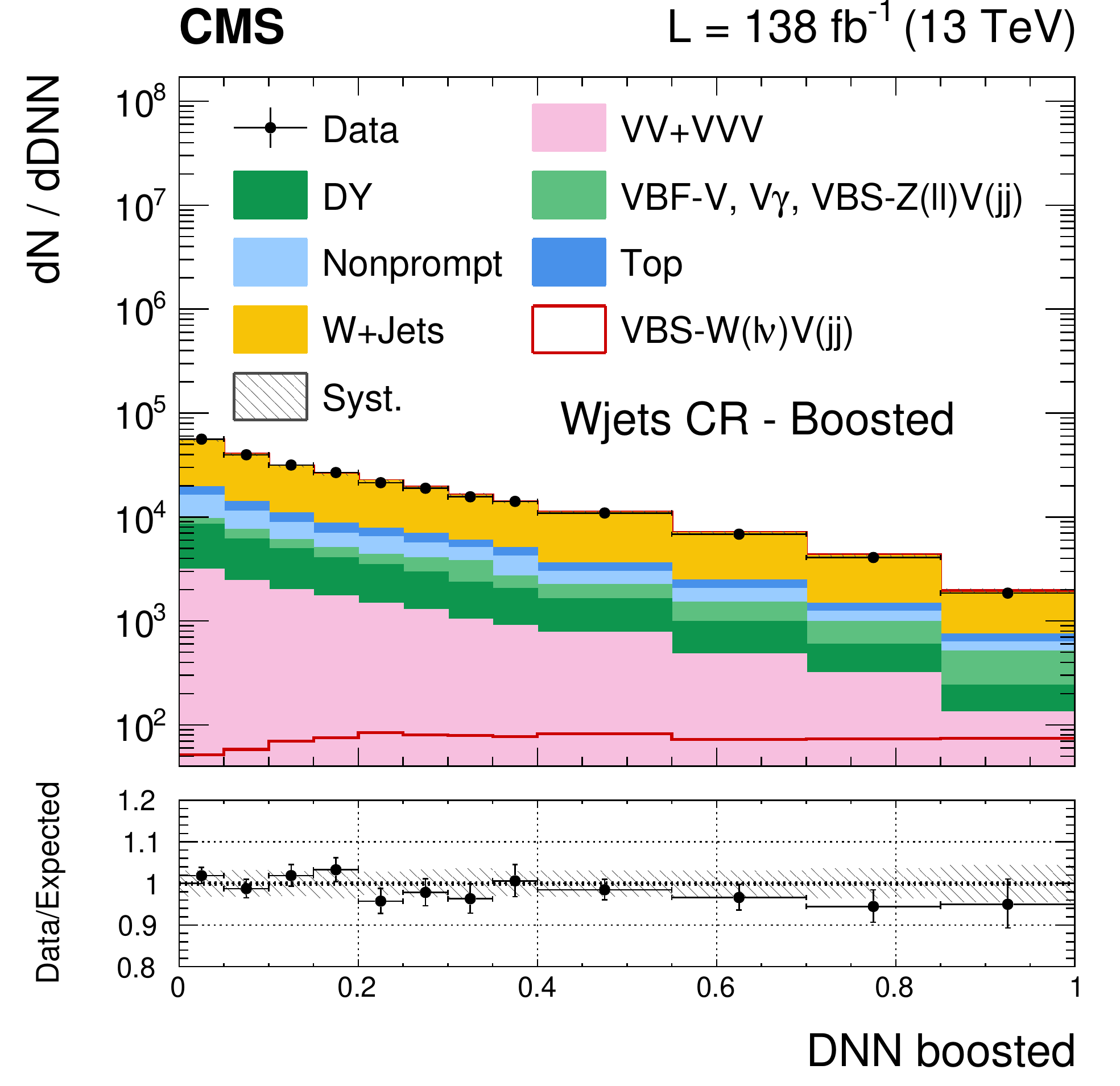}
  \caption{
    The DNN discriminator distribution for the resolved (left) and boosted (right) phase space region in the top quark
    (upper plots) and {\PW}+jets (lower plots) control regions.  Vertical bars on data points show the statistical
    error, whereas the gray band is the post-fit uncertainty on MC with all systematic uncertainties included.
    } \label{fig:CR:dnn}
\end{figure*}

Fig.~\ref{fig:norm:dnnbin} shows the normalized distributions of the DNN discriminator for signal and backgrounds in the
resolved and boosted signal regions.  Fig.~\ref{fig:CR:dnn} shows control plots for the DNN in the top quark and W+jets
control regions both for the resolved and boosted categories.  The predictions and the data agree within the
uncertainties in both cases, after the background estimation based on control samples in data, as described in the
previous section, is applied.

The target of this analysis is the extraction of the $\PW\PV$ VBS production signal strength and the corresponding
significance, obtained through a modified frequentist approach based on the ratio of the experimental likelihood
profiled along with the measurement nuisance parameters over the global likelihood maximum~\cite{Cowan:2010js}.
The likelihood function is built on a signal model taken from simulation and corrected for all residual data/MC
disagreements in particle reconstruction, selection efficiency, and on background models that are built based on
estimates from control samples in data or corrected for data-to-simulation disagreements, as needed.  The DNN
distributions are fitted in the signal phase space regions, combining the two different light lepton flavors, whereas
the yields in the control regions are used only to normalize the {\PW}+jets and top quark backgrounds.

\section{Systematic uncertainties}\label{sec:7}

In the signal extraction fit, each uncertainty is represented by a nuisance parameter that changes the shapes of the
distributions for the signal and background processes or scales their total normalization.  Different sources of
uncertainty are treated as completely uncorrelated in the fit, whereas each uncertainty effect is treated as correlated
or uncorrelated between different channels and processes depending on the case.
In Table~\ref{tab:uncertainty_breakdown} the uncertainties are split into various categories by evaluating the effect on
the total signal strength of freezing each component independently in the fit.

The integrated luminosities of the 2016, 2017, and 2018 data-taking periods are individually known with uncertainties in
the 1.2--2.5\%
range~\cite{CMS-LUM-17-001,CMS-PAS-LUM-17-004,CMS-PAS-LUM-18-002},
whereas the total Run~2 (2016--2018)
integrated luminosity has an uncertainty of 1.6\%.  The improvement in precision reflects the (uncorrelated) time
evolution of some systematic effects.

Discrepancies in the lepton reconstruction and identification efficiencies between data and MC simulation are
compensated by applying correction factors to all samples as functions of the lepton \pt and $\eta$.  Their impact on
the signal region is less than 1\% for both electrons and muons.  The trigger efficiency uncertainty is also smaller
than 1\%.  The electron and muon momentum scale uncertainties are computed by varying the lepton momenta within their
$\pm 1\sigma$ uncertainty, and the resulting uncertainty in the signal yield is less than 1\%. Similarly, jet energy
scale and resolution uncertainties are evaluated by shifting the \pt value of the jets, and thus directly affecting the
reconstructed jet multiplicity and \ptmiss measurement \cite{Khachatryan:2016kdb}; several independent sources are
considered and partially correlated among different data sets, resulting in up to 4\% uncertainty in the signal
strength.

The {\PQb}-tagging data/MC corrections are associated with different uncertainty sources and correlated among all
processes.  Since these uncertainties migrate events between the signal region and top quark control region, they have a
large effect, 5\%, on the signal and background.  Uncertainty in the \ptmiss estimation due to unclustered energy is
also included and calculated by varying the momenta of particles that are not identified with either a jet or a lepton;
its effect is negligible.  Finally, the uncertainty in the pileup modeling is applied to all the relevant MC samples by
varying the minimum bias cross section used to generate the pileup distribution by $\pm 1\sigma$ \cite{CMS:2018mlc} and
estimated to be less than 1\%.

The uncertainty due to the finite number of events in the fitted templates has a significant impact, 10\%, on the signal
measurement. This contribution is dominated by the uncertainty in the nonprompt template estimation, given by the
limited number of data events in the tails and by the large variation of the (positive and negative) weights applied to
model the lepton misidentification probability. Leaving the normalization of the top quark and W+jets (split into many
subcategories) backgrounds floating in the fit results in an uncertainty in the signal strength of 8\%.

The most important theoretical uncertainty is related to the choice of the renormalization and factorization scales in
the MC simulation of events.  The uncertainty in the signal and background yields is computed by taking the largest
variation given by changing such scales up and down independently by a factor of two with respect to their nominal
value, ignoring the extreme case where they are shifted in opposite directions \cite{Cacciari:2003fi,Catani:2003zt}.
The theoretical scale uncertainty is uncorrelated for each background and signal process.  Only shape effects are
included by varying the scales for {\PW}+jets and top quark backgrounds, since their normalization is directly measured
from data in the fit.  Both the shape and normalization effects are included for the other backgrounds.  For the signal,
only the shape effect of the theoretical scales uncertainty is considered while measuring the signal cross section and
significance, whereas the normalization effect is included for the signal strength determination.  Inclusively, the
theoretical scale uncertainty for the EW-only $\PW\PV$ signal is 5\%, and for the QCD-associated diboson production is
25\%.  The overall impact on the EW-only signal strength determination from the choice of renormalization and
factorization scales is 12\%.

The uncertainty in the modeling of the parton shower is also included, by using the weights corresponding to variations
of $\alpS^{\mathrm{ISR}}$ and $\alpS^{\mathrm{FSR}}$ computed by the parton shower programs, and uncorrelated for each
process: the impact on the signal strength determination is 4\%.  The PDF and related strong coupling {\alpS}
uncertainties are evaluated using the eigenvalues of the PDF set following the NNPDF prescription~\cite{Rojo:2016ymp}.
These uncertainties, as well as the one from the modeling of the underlying event, are included for all the processes
apart from top quark and {\PW}+jets backgrounds, and they have a negligible impact on the signal measurement.

\begin{table}[!htb]
  \caption{Breakdown of the uncertainties in the EW $\PW\PV$ VBS signal strength
  measurement.}
  \label{tab:uncertainty_breakdown}
  \centering
  \begin{tabular}{ l c }
  Uncertainty source & $\Delta \mu_{\mathrm{EW}}$ \\
  \hline
  Statistical & 0.12\\
  Limited sample size & 0.10 \\
  Normalization of backgrounds   & 0.08 \\
  Experimental & \\
  \hspace{2em}  {\PQb}-tagging & 0.05 \\
  \hspace{2em}  Jet energy scale and resolution &  0.04 \\
  \hspace{2em} Integrated luminosity & 0.01 \\
  \hspace{2em} Lepton identification & 0.01 \\
  \hspace{2em} Boosted {\PV} boson identification & 0.01 \\
  \hspace{2em} Total & 0.06 \\
  Theory & \\
  \hspace{2em} Signal modeling & 0.09 \\
  \hspace{2em} Background modeling & 0.08 \\
  \hspace{2em} Total & 0.12 \\
  \hline {Total} & 0.22 \\
  \end{tabular}
\end{table}

\section{Results} \label{sec:8}
Three separate maximum likelihood fits are performed: (i) the measurement of the purely EW signal strength $\mew$ keeping
the QCD $\PW\PV$ production contribution fixed to the SM prediction $\mqcd=1$; (ii) the measurement of the signal strength
considering as signal the EW and QCD $\PW\PV$ processes together; (iii) a two-dimensional simultaneous measurement of the
signal strengths $\mew$ and $\mqcd$.

\begin{figure*}[!htb]
  \centering \includegraphics[width=0.49\textwidth]{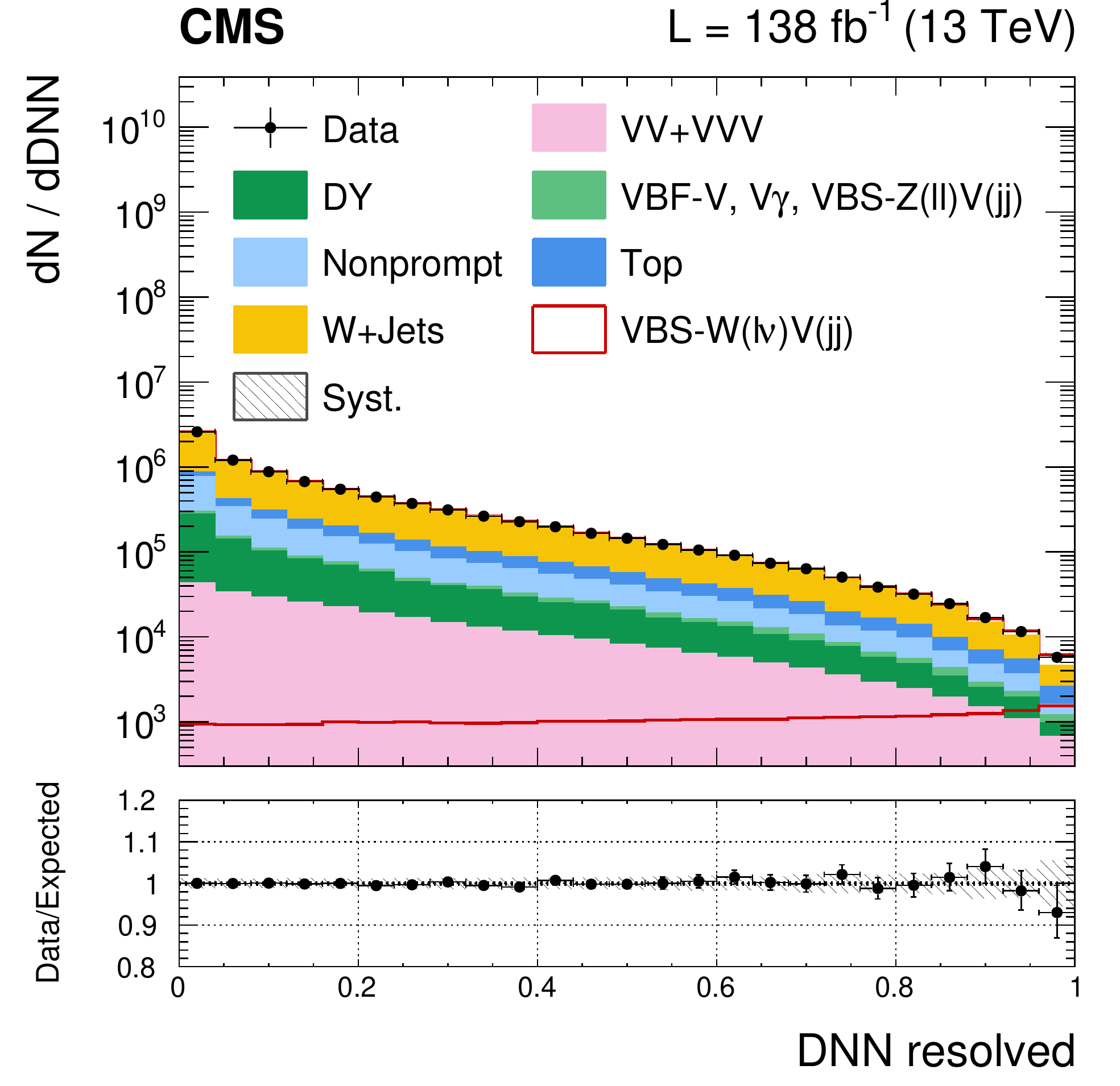} \includegraphics[width=0.49\textwidth]{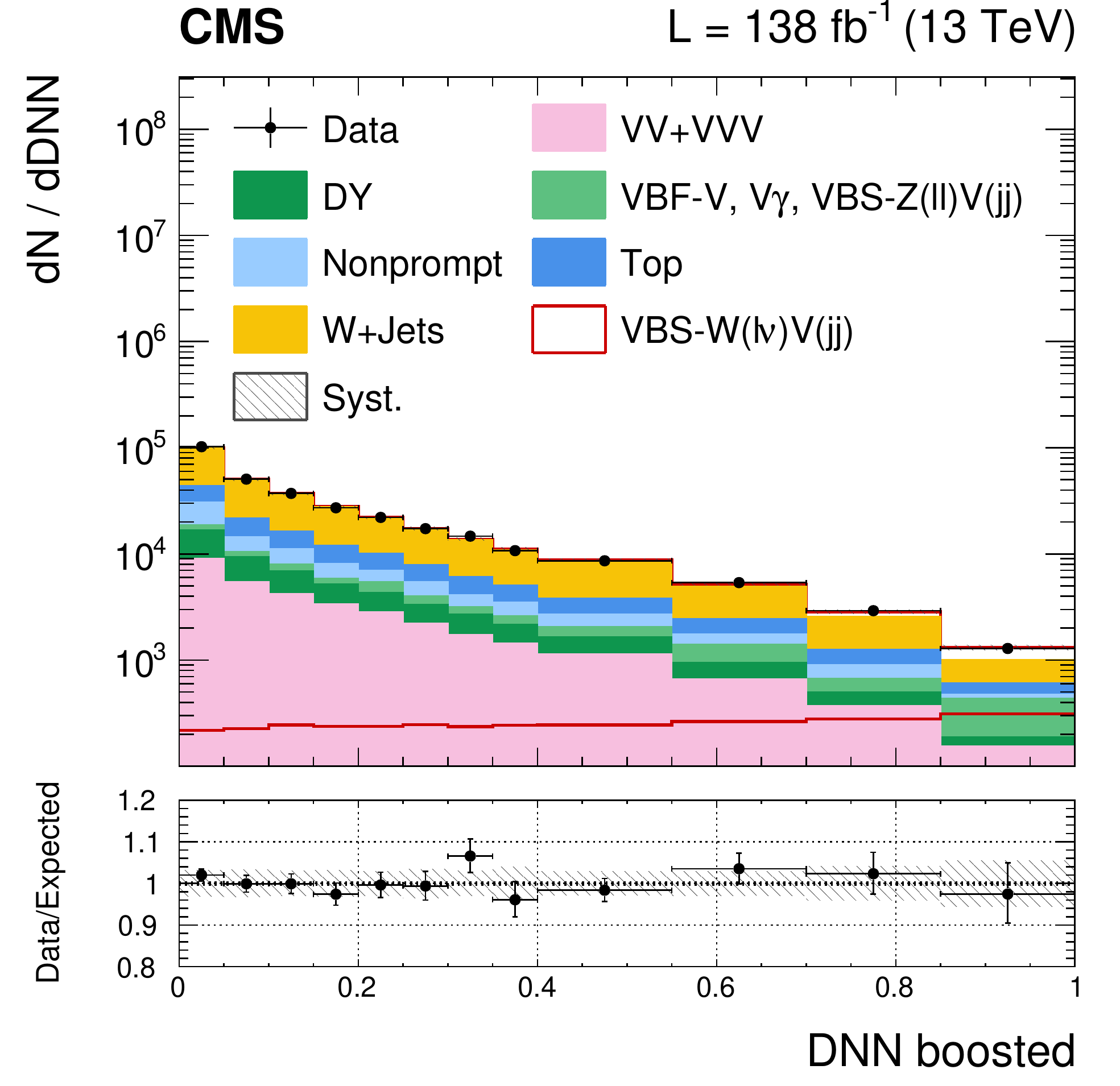} \includegraphics[width=0.49\textwidth]{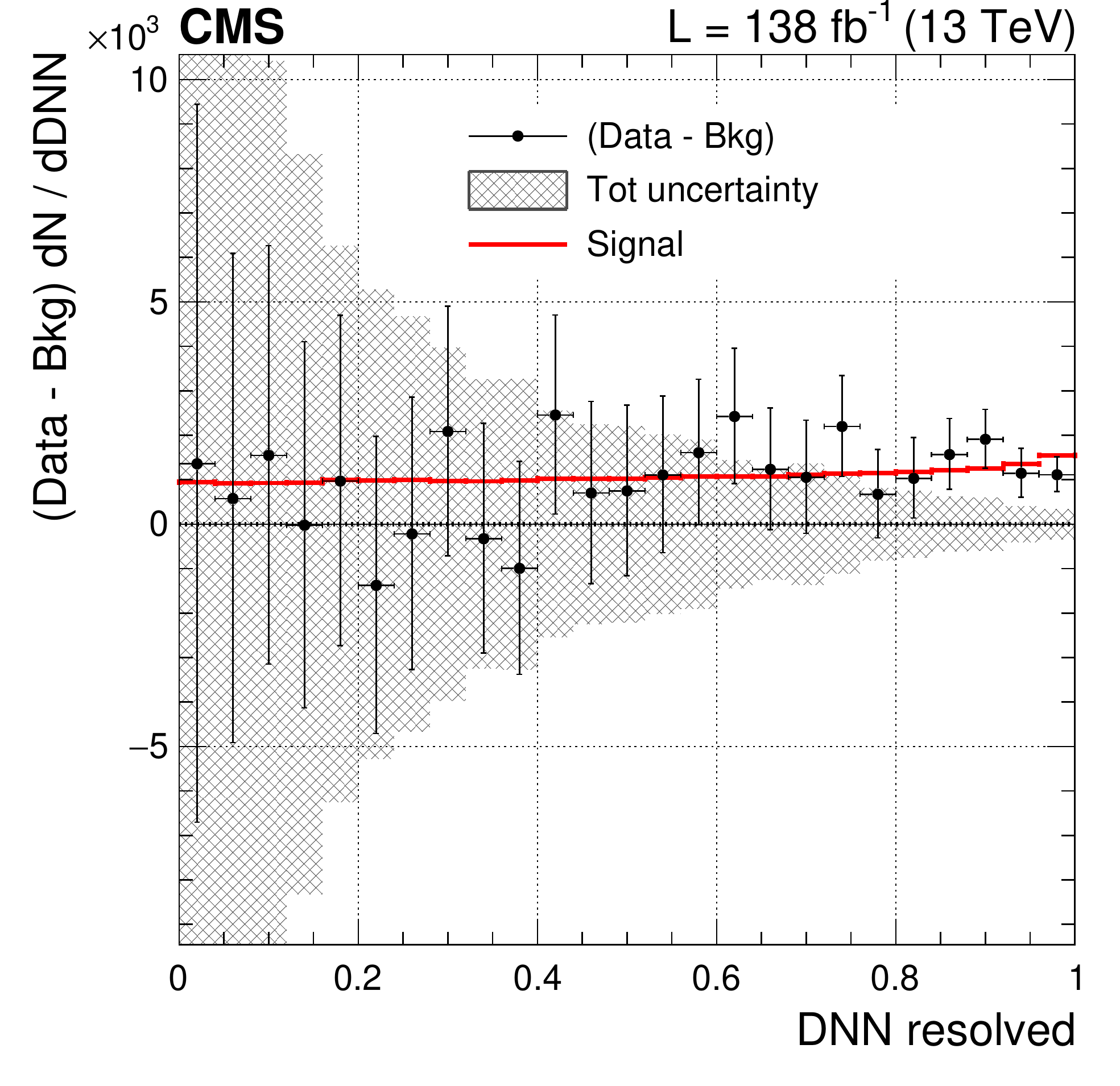} \includegraphics[width=0.49\textwidth]{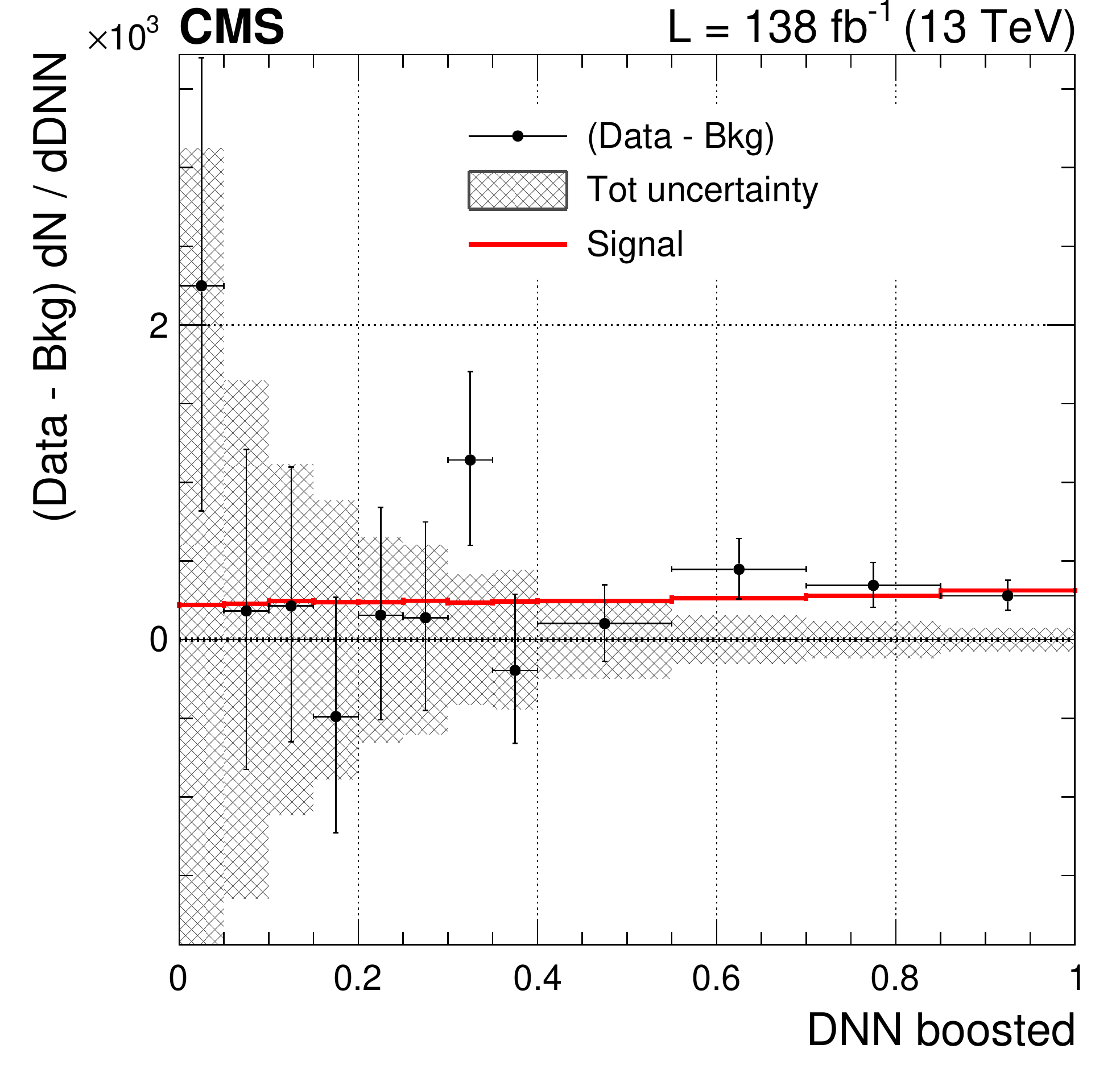} \caption{
    Results for the EW-signal-only fit, keeping the QCD $\PW\PV$ contribution fixed to the SM prediction.  Upper plots:
    post-fit DNN discriminator distributions for the resolved (left) and the boosted (right) signal regions.  The signal
    contribution is plotted both stacked on top of the background processes and also overlaid as a red line to show the signal post-fit
    distribution. The expected yield is the sum of signal and background.  Lower plots: background-subtracted DNN
    discriminator distribution for the resolved (left) and the boosted (right) categories. Post-fit background yields in
    each bin are subtracted from data and compared with the signal post-fit distribution, plotted as a red line.
    Vertical bars on data points show the statistical error, whereas the gray band is the post-fit uncertainty on MC
    with all systematic uncertainties included.  } \label{fig:dnnbin}
\end{figure*}

Figure~\ref{fig:dnnbin} shows the post-fit DNN distribution for the resolved (left) and the boosted (right) signal phase
space in the EW-only fit. The background-subtracted plot, where the evidence for the signal is clearly visible, is also
shown.

A fiducial phase space region is defined at parton level requiring all partons to have $\pt > 10\GeV$ and at least one
pair of outgoing partons with invariant mass $m_{\PQq\PQq}>100\GeV$.  This simple definition of the fiducial region is
chosen for easy application to signal models; the amount of signal generated in this fiducial region and passing
the final signal selection is $\unsim 25\%$.

The SM prediction for the EW $\PW\PV$ production
cross section in this fiducial region is $2.23^{+0.08}_{-0.11}\,(\text{scale}) \pm0.05\,(\text{PDF})$\unit{pb}, where
PDF refers to the uncertainty coming from the parton distribution function.  The measured EW WV production cross section is
$1.90^{+0.53}_{-0.46}$\unit{pb}, corresponding to an observed EW-only signal strength of:
\begin{linenomath}
\begin{equation}
\mew = \frac{\sobs}{\ssm} =  0.85\pm0.12\,(\text{stat})^{+0.19}_{-0.17}\,(\text{syst}) =0.85 ^{+0.23}_{-0.21},
\end{equation}
\end{linenomath}
where \sobs and \ssm are the observed and predicted cross sections, respectively, with an expectation of
$1.00^{+0.24}_{-0.22}$.  The observed significance for the SM EW $\PW\PV$ signal is 4.4 standard deviations with 5.1
expected. The EW-only signal strength fitted independently in the resolved and boosted categories is $0.85\pm0.26$ and
$1.09\pm0.32$, respectively.

\begin{figure}[!htb]
  \centering \includegraphics[width=0.5\textwidth]{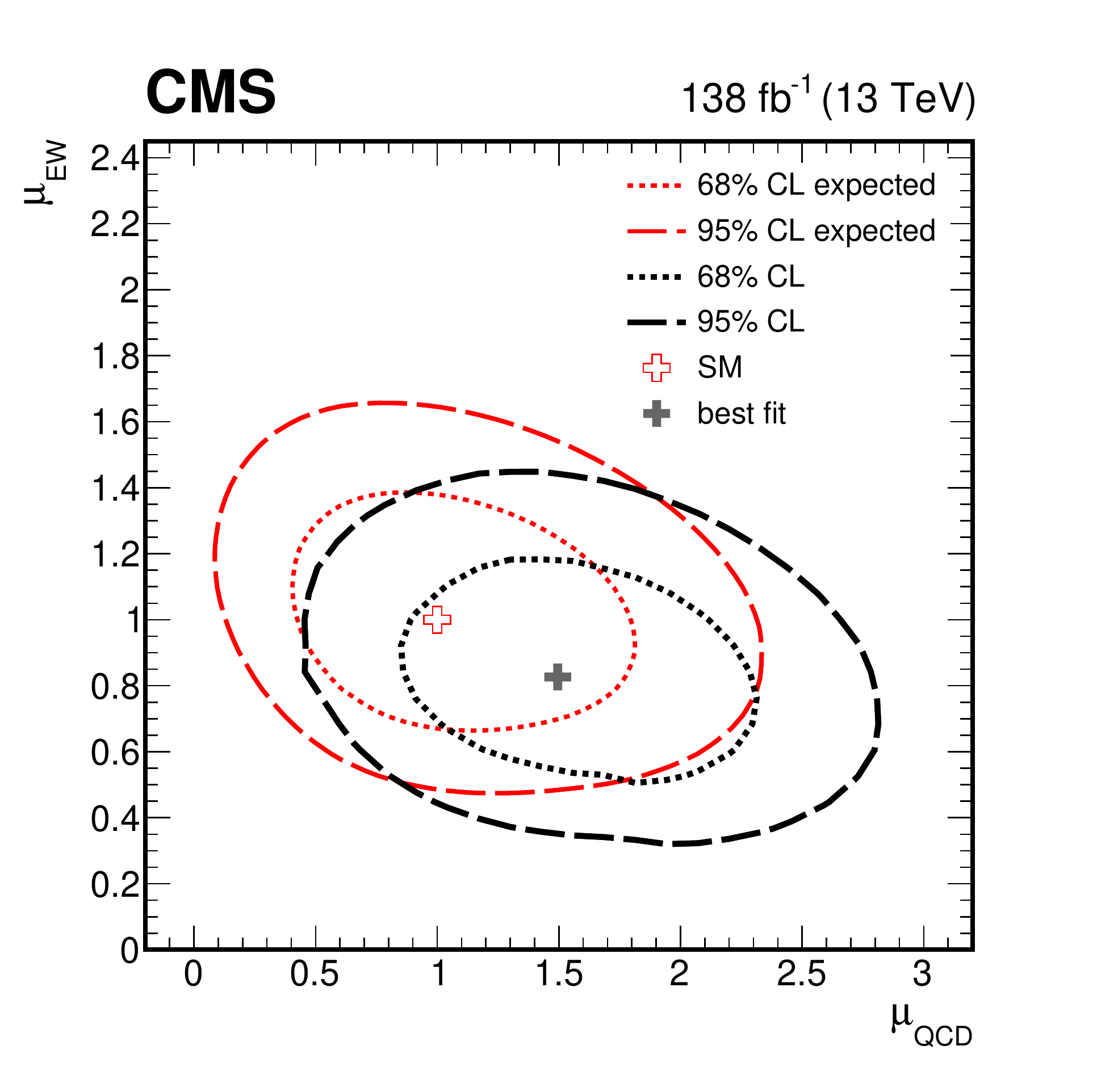} \caption{Simultaneous EW and QCD $\PW\PV$ production
    fit: the expected and observed 68 and 95\% \CL contours on the signal strengths. The best fit result is compatible
    with the SM prediction within the 68\% \CL area.  } \label{fig:fit_2d}
\end{figure}

Considering instead the signal as the overall EW and QCD-associated diboson production, the measured and expected cross
sections are $16.4^{+3.5}_{-2.8}$\unit{pb} and $16.9^{+2.9}_{-2.1}\,(\text{scale}) \pm0.5\,(\text{PDF})$\unit{pb}, respectively,
extracted in the same fiducial phase space region as the EW-only one. This fit assumes the ratio between the EW and
QCD contributions to the diboson production is fixed to the value predicted by the SM.  The overall signal strength
$\mu = \sobs / \ssm$, with an expectation of $1.00^{+0.21}_{-0.20}$, is measured as:
\begin{linenomath}
\begin{equation}
\mu_{\mathrm{EW}+\mathrm{QCD}} =  0.97\pm0.06\,(\text{stat})^{+0.19}_{-0.21}\,(\text{syst})= 0.97 ^{+0.20}_{-0.22} .
\end{equation}
\end{linenomath}
The fit is also performed leaving as free independent parameters the signal strengths of the EW and QCD-associated
$\PW\PV$ production components ($\mew$ and $\mqcd$).  The result of the 2D fit is shown in Fig.~\ref{fig:fit_2d}, where
the expected and observed minima are presented, together with the 68 and 95\% confidence level (\CL) contours built from
the likelihood function.  The measured signal strengths are in agreement with the SM predictions within the 68\% \CL.

\section{Summary}\label{sec:9}

The first evidence for the electroweak (EW) production of a $\PW\PV$ ($\PV = \PW$ or {\PZ}) pair plus two jets in the
$\ell\nu\PQq\PQq$ decay channel is reported.  Events are separated into two categories: either the hadronically decaying
{\PW} or {\PZ} boson is reconstructed as one large-radius jet, or it is identified as a pair of jets with dijet mass
close to the boson mass.  Multivariate machine learning discriminators are optimized to separate the signal from the
background in each category and their outputs are exploited in the statistical analysis.  The large background from
single {\PW} boson production accompanied by jets is estimated from control samples in the data to reduce the impact of
Monte Carlo mismodeling in this multijet phase space region.

Tabulated results are provided in the HEPData record for this analysis~\cite{HEPData}.

The EW-only $\PW\PV$ signal strength, measured keeping the QCD-associated diboson production fixed to the standard model
prediction, is: $\mew = \sobs / \ssm = 0.85\pm0.12\,(\text{stat})^{+0.19}_{-0.17}\,(\text{syst}) = 0.85 ^{+0.23}_{-0.21}
$ at $1.00^{+0.24}_{-0.22}$ expected, where \sobs and \ssm are the observed and predicted cross sections, respectively.
The observed significance for the SM EW $\PW\PV$ signal is 4.4 standard deviations with 5.1 expected., it is measured
keeping the quantum chromodynamics (QCD) associated diboson production fixed to the standard model prediction.  When we
consider the signal as the total EW and QCD-associated diboson yield, the overall signal strength
$\mu_{\mathrm{EW}+\mathrm{QCD}}$ is measured as: $0.97\pm0.06\,(\text{stat})^{+0.19}_{-0.21}\,(\text{syst}) = 0.97
^{+0.20}_{-0.22}$ with an expectation of $1.00^{+0.21}_{-0.20}$.  Finally, a simultaneous two-dimensional fit of the EW
and QCD $\PW\PV$ production components is performed.

Overall, both the $\PW\PV$ EW-only measurement and the simultaneous EW and QCD $\PW\PV$ measurements are in agreement
with the SM predictions within the 68\% confidence level.

\begin{acknowledgments}
  We congratulate our colleagues in the CERN accelerator departments for the excellent performance of the LHC and thank   the technical and administrative staffs at CERN and at other CMS institutes for their contributions to the success of   the CMS effort. In addition, we gratefully acknowledge the computing centers and personnel of the Worldwide LHC   Computing Grid and other centers for delivering so effectively the computing infrastructure essential to our   analyses. Finally, we acknowledge the enduring support for the construction and operation of the LHC, the CMS   detector, and the supporting computing infrastructure provided by the following funding agencies: BMBWF and FWF   (Austria); FNRS and FWO (Belgium); CNPq, CAPES, FAPERJ, FAPERGS, and FAPESP (Brazil); MES and BNSF (Bulgaria); CERN;   CAS, MoST, and NSFC (China); MINCIENCIAS (Colombia); MSES and CSF (Croatia); RIF (Cyprus); SENESCYT (Ecuador); MoER,   ERC PUT and ERDF (Estonia); Academy of Finland, MEC, and HIP (Finland); CEA and CNRS/IN2P3 (France); BMBF, DFG, and   HGF (Germany); GSRI (Greece); NKFIA (Hungary); DAE and DST (India); IPM (Iran); SFI (Ireland); INFN (Italy); MSIP and   NRF (Republic of Korea); MES (Latvia); LAS (Lithuania); MOE and UM (Malaysia); BUAP, CINVESTAV, CONACYT, LNS, SEP, and   UASLP-FAI (Mexico); MOS (Montenegro); MBIE (New Zealand); PAEC (Pakistan); MSHE and NSC (Poland); FCT (Portugal); JINR   (Dubna); MON, RosAtom, RAS, RFBR, and NRC KI (Russia); MESTD (Serbia); MCIN/AE and PCTI (Spain); MOSTR (Sri Lanka);   Swiss Funding Agencies (Switzerland); MST (Taipei); ThEPCenter, IPST, STAR, and NSTDA (Thailand); TUBITAK and TAEK   (Turkey); NASU (Ukraine); STFC (United Kingdom); DOE and NSF (USA).

 \hyphenation{Rachada-pisek} Individuals have received support from the Marie-Curie program and the European Research  Council and Horizon 2020 Grant, contract Nos.\ 675440, 724704, 752730, 758316, 765710, 824093, 884104, and COST Action  CA16108 (European Union); the Leventis Foundation; the Alfred P.\ Sloan Foundation; the Alexander von Humboldt  Foundation; the Belgian Federal Science Policy Office; the Fonds pour la Formation \`a la Recherche dans l'Industrie et  dans l'Agriculture (FRIA-Belgium); the Agentschap voor Innovatie door Wetenschap en Technologie (IWT-Belgium); the  F.R.S.-FNRS and FWO (Belgium) under the ``Excellence of Science -- EOS" -- be.h project n.\ 30820817; the Beijing  Municipal Science \& Technology Commission, No. Z191100007219010; the Ministry of Education, Youth and Sports (MEYS) of  the Czech Republic; the Deutsche Forschungsgemeinschaft (DFG), under Germany's Excellence Strategy -- EXC 2121  ``Quantum Universe" -- 390833306, and under project number 400140256 - GRK2497; the Lend\"ulet (``Momentum") Prog ram  and the J\'anos Bolyai Research Scholarship of the Hungarian Academy of Sciences, the New National Excellence  Program \'UNKP, the NKFIA research grants 123842, 123959, 124845, 124850, 125105, 128713, 128786, and 129058 (Hungary);  the Council of Science and Industrial Research, India; the Latvian Council of Science; the Ministry of Science and  Higher Education and the National Science Center, contracts Opus 2014/15/B/ST2/03998 and 2015/19/B/ST2/02861 (Poland);  the Funda\c{c}\~ao para a Ci\^encia e a Tecnologia, grant CEECIND/01334/2018 (Portugal); the National Priorities  Research Program by Qatar National Research Fund; the Ministry of Science and Higher Education, projects  no. 14.W03.31.0026 and no. FSWW-2020-0008, and the Russian Foundation for Basic Research, project No.19-42-703014  (Russia); MCIN/AEI/10.13039/501100011033, ERDF ``a way of making Europe", and the Programa Estatal de Fomento de la  Investigaci{\'o}n Cient{\'i}fica y T{\'e}cnica de Excelencia Mar\'{\i}a de Maeztu, grant MDM-2017-0765 and Programa  Severo Ochoa del Principado de Asturias (Spain); the Stavros Niarchos Foundation (Greece); the Rachadapisek Sompot Fund  for Postdoctoral Fellowship, Chulalongkorn University and the Chulalongkorn Academic into Its 2nd Century Project  Advancement Project (Thailand); the Kavli Foundation; the Nvidia Corporation; the SuperMicro Corporation; the Welch  Foundation, contract C-1845; and the Weston Havens Foundation (USA).
\end{acknowledgments}
 
\bibliography{auto_generated}

\cleardoublepage \appendix\section{The CMS Collaboration \label{app:collab}}\begin{sloppypar}\hyphenpenalty=5000\widowpenalty=500\clubpenalty=5000\cmsinstitute{Yerevan~Physics~Institute, Yerevan, Armenia}
A.~Tumasyan
\cmsinstitute{Institut~f\"{u}r~Hochenergiephysik, Vienna, Austria}
W.~Adam\cmsorcid{0000-0001-9099-4341}, J.W.~Andrejkovic, T.~Bergauer\cmsorcid{0000-0002-5786-0293}, S.~Chatterjee\cmsorcid{0000-0003-2660-0349}, K.~Damanakis, M.~Dragicevic\cmsorcid{0000-0003-1967-6783}, A.~Escalante~Del~Valle\cmsorcid{0000-0002-9702-6359}, R.~Fr\"{u}hwirth\cmsAuthorMark{1}, M.~Jeitler\cmsAuthorMark{1}\cmsorcid{0000-0002-5141-9560}, N.~Krammer, L.~Lechner\cmsorcid{0000-0002-3065-1141}, D.~Liko, I.~Mikulec, P.~Paulitsch, F.M.~Pitters, J.~Schieck\cmsAuthorMark{1}\cmsorcid{0000-0002-1058-8093}, R.~Sch\"{o}fbeck\cmsorcid{0000-0002-2332-8784}, D.~Schwarz, S.~Templ\cmsorcid{0000-0003-3137-5692}, W.~Waltenberger\cmsorcid{0000-0002-6215-7228}, C.-E.~Wulz\cmsAuthorMark{1}\cmsorcid{0000-0001-9226-5812}
% ...
\vskip\cmsinstskip
% ...
1:~Also at TU Wien, Wien, Austria\\
% ...
\end{sloppypar}
\end{document}